\documentclass[pdflatex,sn-mathphys-num]{sn-jnl}

\usepackage{graphicx}%
\usepackage{multirow}%
\usepackage{amsmath,amssymb,amsfonts}%
\usepackage{amsthm}%
\usepackage{mathrsfs}%
\usepackage[title]{appendix}%
\usepackage{xcolor}%
\usepackage{textcomp}%
\usepackage{manyfoot}%
\usepackage{booktabs}%
\usepackage{algorithm}%
\usepackage{algorithmicx}%
\usepackage{algpseudocode}%
\usepackage{listings}%
\usepackage{enumitem}

\usepackage[capitalise,noabbrev]{cleveref}

\usepackage[range-phrase=\text{ to },per-mode=symbol-or-fraction,binary-units=true,range-units=single,list-units=single,detect-all]{siunitx}

\usepackage{acro}
\DeclareAcronym{STEM}{short=STEM, long={Science, Technology, Engineering and Mathematics}}
\DeclareAcronym{ICT}{short=ICT, long=Information and Communication Technology}
\DeclareAcronym{MINT}{short=MINT, long={Mathematics, Information technology, Natural sciences and Technology}}
\DeclareAcronym{CS}{short=CS, long={computer science}}

\usepackage{lineno}

\begin{document}

\title[Building a Bridge between the Two Schools]{\huge \bfseries Building a Bridge between the Two Schools: Realizing a Practical Path to Include Literacy-based Skills within the STEM Curricula} 


\author*[1]{\fnm{Jorge} \sur{Torres Gómez}}\email{jorge.torresgomez@ieee.org}

\author[2]{\fnm{Erika} \sur{Gericke}}\email{erika.gericke@uni-siegen.de}
\equalcont{These authors contributed equally to this work.}

\author[3]{\fnm{Anton} \sur{Rass\~{o}lkin}}\email{anton.rassolkin@taltech.ee}
\equalcont{These authors contributed equally to this work.}

\author[4]{\fnm{Mikołaj} \sur{Leszczuk}}\email{mikolaj.leszczuk@agh.edu.pl}
\equalcont{These authors contributed equally to this work.}

\author[5]{\fnm{Alexandru} \sur{Iosup}}\email{A.Iosup@vu.nl}
\equalcont{These authors contributed equally to this work.}

\author[6]{\fnm{Marcin} \sur{Niemiec}}\email{marcin.niemiec@agh.edu.pl}
\equalcont{These authors contributed equally to this work.}

\author[7]{\fnm{Carmen} \sur{Peláez-Moreno}}\email{carmen@tsc.uc3m.es}
\equalcont{These authors contributed equally to this work.}

\affil*[1]{\orgname{TU Berlin}, \orgaddress{\street{Einstenuferstr. 25, FT5}, \city{Berlin}, \postcode{10587}, \country{Germany}}}

\affil[2]{\orgname{University of Siegen}, \orgaddress{\street{Adolf-Reichwein-Straße 2a}, \city{Siegen}, \postcode{57076}, \country{Germany}}}

\affil[3]{\orgname{Tallinn University of Technology}, \orgaddress{\street{Ehitajate tee 5}, \city{Tallin}, \postcode{19086}, \country{Estonia}}}

\affil[4,6]{\orgname{AGH University of Kraków}, \orgaddress{\street{ al. Adama Mickiewicza 30}, \city{Kraków}, \postcode{30-059}, \country{Poland}}}

\affil[5]{\orgname{Vrije Universiteit Amsterdam}, \orgaddress{\street{De Boelelaan 1105}, \city{Amsterdam}, \postcode{1081}, \country{The Netherlands}}}

\affil[7]{\orgname{Universidad Carlos III de Madrid, UC3M}, \orgaddress{\street{Av. Universidad, 30}, \city{Madrid}, \postcode{28911}, \country{Spain}}}



\abstract{
Developing students as well-rounded professionals is increasingly important for our modern society.
Although there is a great consensus that technical and professional ("\textit{soft}") skills should be developed and intertwined in the core of computer science subjects, there are still few examples of alike teaching methodologies at technical schools.
This contribution investigates the integration of technical and professional skills while teaching specialized curricula in computer science.
We propose a broadly applicable, step-by-step methodology that connects core technical concepts (e.g., information entropy, network security) with fine arts practices such as music, video production, gaming, and performing arts (e.g., Oxford-style debates).
The methodology was applied in several computer science courses at technical universities, where quantitative and qualitative assessments, including student questionnaires and exam scores, showed improved learning outcomes and increased student engagement compared to traditional methods.
The results indicate that this art-based integration can effectively bridge the historical divide between the two schools of thought, offering a practical direction for educators.
Within this context, we also identify open issues that will guide future research on topics such as instructor engagement, female motivation in technical subjects, and scalability of these approaches.
}


\maketitle

\section{Introduction}

``Is it sufficient to teach only \textit{hard concepts} in STEM\footnote{\ac{STEM}} curricula?'', 
``How to educate with broader skills beyond the technical ones?''
Answering such questions is essential in setting a curricular direction, and can mean the difference between developing students exclusively into technical experts or well-rounded professionals \cite{walker2010computer}. 
In this work, we take the standpoint that the latter is more desirable for both the current job market and our society. 
Lacking today an effective integration of \textit{professional skills}\footnote{We use \textit{professional skills} in the sense of the ones supporting social interactions (typically named as \textit{soft skills}) in contrast to hard skills in the sense of \textit{technology skills} \cite{berdanier2022hard}.} in \ac{STEM} curriculum, we focus on answering the key question ``How to naturally teach the professional skills at technical classes?'' and propose to do so through fine art-based methods.

As stated in Snow's speech \cite{snow_two_2013}, contemporary and modern academic programmes have historically been conceived in two separate schools regarding the arts and natural sciences fields. 
Education in the arts is commonly associated with disciplines such as music, performance, and painting, whereas the \ac{STEM} fields are more focused on mathematics, science, and technology.
Historically, an \textit{artificial} barrier has emerged, leading to two different cultures, intellectuals who are literate and natural scientists.

This question naturally arises with the convergence of the two traditionally separate schools in liberal arts curricula \cite{consortium2007model}.
More recently, within the research community, there is a remarkable consensus on a broader understanding of the teaching of professional skills in technical schools toward the development of professional skills~\cite{walker2010computer,caratozzolo2022use,aizenman2024liberal}.
A tender ``academic teaching movement'' is slowly coming to the fore-- university educators focusing on the requirement to teach their students much-needed professional skills.
The work related to the development of new teaching methodologies and practices, such as those based on fine arts approaches, is becoming visible; see, for instance, the discussion in \cite{repenning2020computational,pyshkin2017liberal,barker2005what} and examples of interdisciplinary curricula in computer science in liberal arts colleges in \cite{teresco2022cs,hollandminkley2023computer}.

\begin{figure}
\centering
\includegraphics[width=0.5\textwidth]{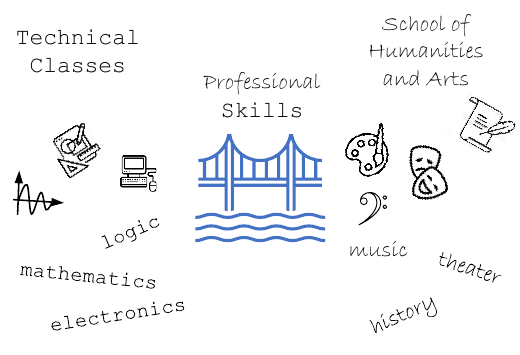}
\caption{Bridging the two schools towards reinforcing professional skills at technical subjects.}
\label{fig:intro}
\end{figure}

Although liberal arts colleges implement mixed classes to teach professional and technical skills (see examples in~\cite[Sec. 2]{walker2010computer}, \cite{aizenman2024liberal}), a separation still exists.
The liberal arts curriculum allows students to attend courses outside the major (around $\SIrange{40}{50}{\percent}$ of the curriculum) while pursuing a major in \ac{CS}, and tacitly, it is assumed that students will apply a broader view for addressing and solving technical problems with the perspectives learned within humanity classes.
In practice, the in-place technical classes do not implement methodologies from humanity subjects by design, which is an impediment towards the conception of liberal arts.
The liberal arts curriculum design mostly relies on senior project classes to develop professional skills within a technical class; see \cite[Sec. 3.1]{bruce2010history} and case studies in \cite{baldwin2010case}.
In our view, teaching technical classes in \ac{STEM} majors through methodologies that include fine-arts practices, or even the direct use of fine-arts methodologies, can further blur these intellectual boundaries and break this historical construct of two different classes.

The deficiency in teaching professional skills is more markedly at technical universities, where 
daily practice limits content beyond exclusively technical matters~\cite{kamp2016engineering}. 
Challenges include
(1) the overcrowded curriculum due to the ever-increasing presence of new technological paradigms (such as Industry 4.0, cloud computing and cyber security, 5G and 6G communications, etc.),
(2) the complexity of technical concepts to teach (i.e., in artificial intelligence and machine learning, in computer systems and networks, in data science, etc.) and
(3) the limited time to develop new teaching activities. 

In general, intertwining fine arts methodologies in the teaching of technical subjects could directly address broader abilities within the same class, such as critical thinking, teamwork, or oral presentation skills.
Also, not only students but educators would benefit: fine arts-based approaches could provide a natural way to incorporate the teaching of broader skills into technical classes, lessening hurdles for educators.
Naturally, the question arises when considering overcoming these challenges in daily practice: ``How can these two schools effectively integrate to support the teaching of professional skills in technical classes?''
In this paper, we illustrate a pragmatic direction to answer this question.
We propose a methodology as a backbone for integrating professional skills into technical classes.
We also provide illustrative examples of applications in different computer science technical subjects based on our experience at various institutions.

\section{Technical Box: The Developing State of Teaching Professional Skills}
\label{sec:related_work}

Some curricula already focused on covering broader skills through the addition of arts-related courses, but separately from the main curriculum, see, for instance, the programmes implemented at technical universites in~\cite{noauthor_pasaporte_nodate,noauthor_moses_nodate,noauthor_openspace_nodate}.
There, students focus exclusively on writing, communication, or teamwork collaboration.

Groundbreaking work also elaborates on the integration of the fine arts into the design of specific teaching activities for \ac{STEM} classes.
For example,
(1) student motivation can increase when arts and animation are integrated with computer programming subjects~\cite{jawad_integrating_2018,rebelsky2013building}, 
(2) student learning can improve when music, 3D art, and emoticons appear in programming, algebra and geometry classes \cite{barmpoutis_integrating_2018,weiss2018perfect}, 
(3) students experience a more tangible experience with animations that can potentially improve and enrich their understanding of abstract mathematical objects~\cite{abu_mathematical_2001}, and
(4) social concerns such as genre disbalance or historical events can be grounded when designing assignments for coding \cite{lionelle2020cs}.  
All of these are principled and well-tested methodologies to motivate and engage students in the learning of difficult topics.

However, such ad-hoc approaches do not test how students integrate professional and technical skills.
Experience in the field indicates that students and educators struggle to bring over professional skills~\cite{barak2022conceptualization}, e.g., when trying to report on their final undergraduate or graduate projects, when working as teams on goal-orientated projects and confronting their ideas with those of others, when facing ethical and social consequences of technological developments, when realizing about the real-world and social constraints, etc.

Then, how can these particular experiences be generalized for broader applicability and scope?
We pose the need for a methodology that systematically integrates fine-arts practices into the \ac{STEM} learning so that they become the backbone of acquiring professional and technical skills together.

%

\section{Bridging the Two Schools}
\label{sec:methods}

\begin{figure}
\centering
\includegraphics[width=0.6\columnwidth]{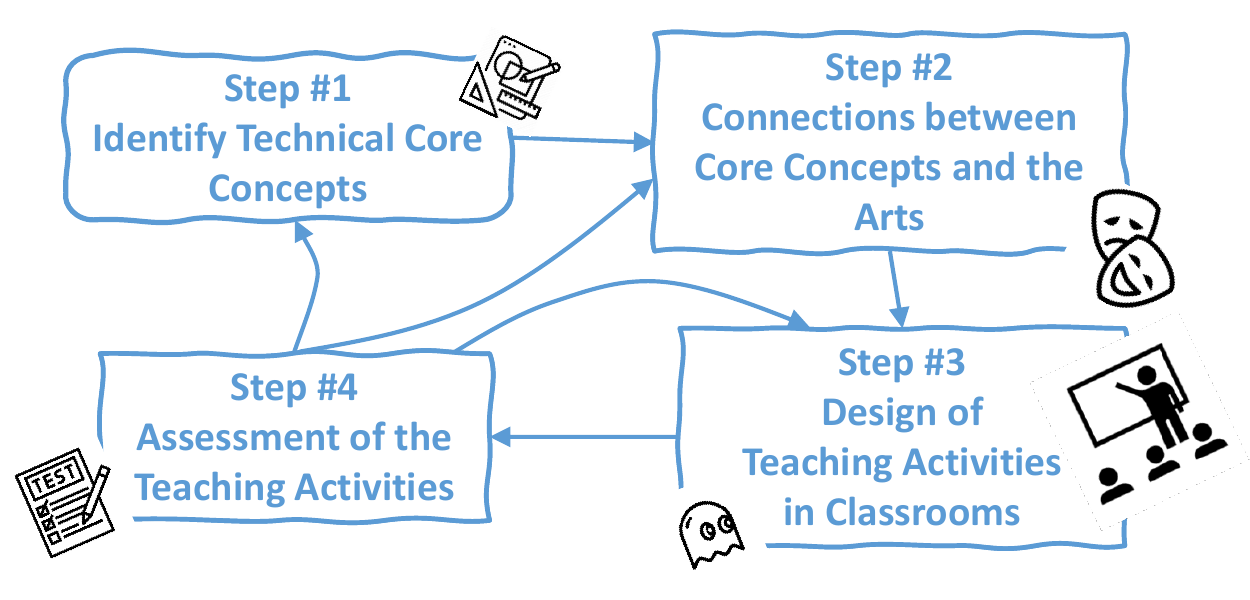} 
\caption{A step-by-step methodology to bridge the two schools.}
\label{fig:methodology}
\end{figure}

Here we further elaborate on a set of step-by-step guidelines to facilitate the inclusion of fine arts methodologies into \ac{STEM} topics.
Through the core methodology illustrated in \Cref{fig:methodology}, we aim to enrich the learning of \textit{technical topics} by integrating professional skills through the arts, rather than by adding separate humanity-related subjects to the curriculum as an ``additional''.

\begin{enumerate}[label=\textbf{\textit{Step~\arabic*.}}, labelindent=0pt, wide, labelwidth=!]

    \item \textbf{To Identify the Technical Core Concepts} of the subject to be enriched. For this methodology to be effective, these concepts must truly be the basis for the subject.
    This requires experienced instructors to detect and identify key aspects that facilitate the understanding of technical concepts. 
    This could be carried out by undertaking a hierarchical analysis of the competencies required from the students or by an ad hoc critical review of the most common errors observed in previous editions of the course. 
    \item \textbf{To Find Connections between Core Concepts and the Arts}
    by establishing bonds between the previously identified core concepts and any form or expression of arts. 
    Here, we would like to underline that arts and \ac{STEM} are complex per se and the connections we should find will be necessarily nuanced. 
    We encourage instructors to creatively explore the affordances of arts to communicate and engage beyond those already existing connections between technology and arts (e.g. digital design, artificial intelligence generated art like current diffusion models, robotic art, etc.)
    The connections in this step should be in the service of communicating the nuclear ideas selected independently in the first step, avoiding shallower or easier-to-find metaphors with other secondary concepts.
    \item \textbf{To design and implement teaching activities}
    based on the connections identified in step two (illustrative examples of these activities will be provided in \Cref{sec:examples}). 
    A set of requisites of these activities should be enforced such as fostering critical thinking, creativity, active learning, or teamwork.
    This allows diverse ways of participation and diverse types of solutions when considering the broad inclusiveness of all of the students' profiles.
    We propose using references and contexts relevant to the students and their future profession where they are encouraged to express themselves and become aware of the importance of different points of view.
    Specifically, rigid templates that do not foster rhizomatic learning should be avoided. 
    \item \textbf{To Assess the Results of the Implementation of the Teaching Activities}
    in different stages and several dimensions. This last step allows us to improve the previous three steps in future iterations.
    From the point of view of temporal sequencing, assessment should take place 
    before, during, and after the course. 
    The first is intended to collect the students' points of view and satisfaction, including their suggestions. This facilitates the discovery of meaningful relationships and minimises the often present mismatches between the students and the instructors' cultural and artistic references. Likewise, the second is designed to understand the instructors' opinions, and finally the third evaluates the knowledge and competencies acquired by the students.    
\end{enumerate}

%

\section{Direct Experiences in the Classroom}
\label{sec:examples}

Including arts methodologies in daily practice in technical courses may be more straightforward than expected.
In particular, those hard-to-follow technical concepts are reliably learned when building on their qualitative description provided by arts and literacy practices.
Reflecting on concepts through the arts, performing or gaming inherently provides a platform to naturally mix both schools.
The teaching of technical concepts can be conceived with \textit{less-conventional} methods directly stimulating critical thinking, motivation, and social engagement.
Based on our personal experience in classrooms, we introduce the application of the above methodology to the computer science-related curriculum, and, in the text, we signal its applicability with parentheses.

%

\subsection{Intertwining Art and Literacy with Communication Theory}

Arts and literacy provide means to teach technical curricula with a multisensorial approach.
Appreciating and producing art, for instance, when painting or producing music, and including writing essays and poems, are practices from liberal arts methodologies that stimulate creativity and curiosity on their own.
To illustrate the use of arts and literacy to teach communication theory, we target the concept of information entropy, as it plays a central role in explaining the coding and decoding of information (\textit{Step~\num{1}}).
Although it is defined analytically, its comprehension may be reached through its qualitative representation in terms of complexity.
In this way, we can analyse and produce pieces of art, essays, video, or music records concerning its complexity using the concept of entropy (\textit{Step~\num{2}}) \cite{torres-gomez2021teaching}. 
Due to its manifest practice, teaching technical concepts through the methodologies of the liberal arts inherently provides a more tangible experience of the usual analytical definition of those.

\begin{figure}[!htb]
	\centering
	\includegraphics[width=0.7\columnwidth]{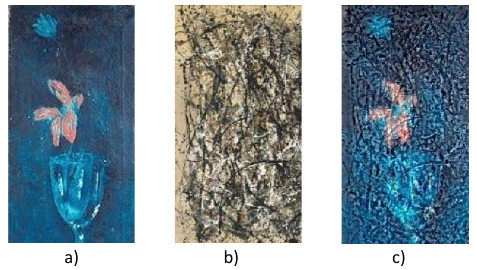}
	\caption{Illustrative example of painters and mixed styles. a) Picasso, “The Blue Cup”, 1902. b) Pollock, “Number 31”, 1950. c) The mixed style of Pollock and Picasso \cite{torres-gomez2021teaching}.}
	\label{fig:methodology_1}
\end{figure}

As an example with images, we illustrate complexity with the fractal style of Jackson Pollock \cite{taylorPerceptualPhysiologicalResponses2011} in contrast to the blue period frames of Pablo Picasso (cf. \Cref{fig:methodology_1}). 
Students are asked to measure and compare both images' complexity or chaoticity using the analytic definition of information entropy (\textit{Step~\num{3}}).
This practice provides a more tangible experience of this concept when reflecting the diversity of pictorial information.\footnote{See an example of pictorial information relevancy in \cite{frank2021state}.}
Exercises may also be conducted to synthesize, not only to analyze.
Students are asked to produce a new frame by mixing the entropy levels of both frames, which results in the combined image in \Cref{fig:methodology_1} c).
Both styles are mixed to introduce some chaotic behavior in the Picasso frame according to the measured entropy levels of Pollock.\footnote{Further details of this procedure are exposed in \cite{torres_gomez_teaching_2021}.}

As another example with the written text~\cite{torres-gomez2020towards}, the concept of entropy can be used to evaluate the level of redundancy in poems in connection with the perception of rhythms and metrics in different periods, i.e. \textit{Renaissance} and \textit{Contemporary} (\textit{Step~2}).
Exercises may be conducted to reduce redundancy by rephrasing the verse's prose while preserving its content's meaning and beauty (\textit{Step 3}).
In this way, we can connect qualitative perception and the quantitative nature of entropy.\footnote{Interested readers may review the poems collection by Taklaja \cite{Taklaja2016}, as written by students at (TalTech University, Estonia).}

\begin{figure}
    \centering
    \includegraphics[width=0.7\columnwidth]{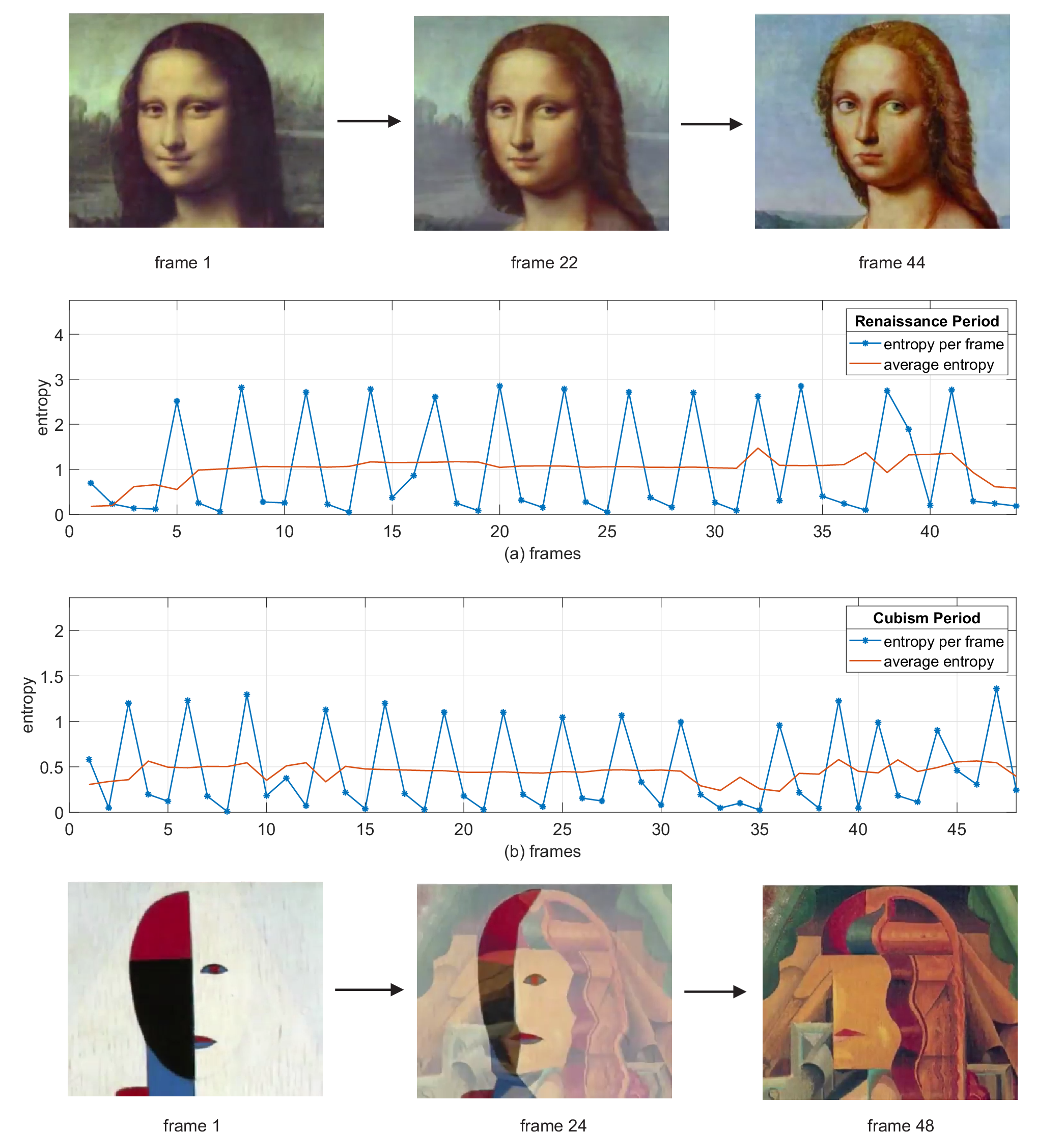}
    \caption{Measuring \textit{entropy} in two video sequences from  "500 Years of Women in Western Art" by Philip Scott Johnson \cite{torres-gomez2020teaching}.}    
\label{fig:entropy}
\end{figure}

Furthermore, using illustrative video sequences, entropy may be used to evaluate the chaoticity of a single frame over time; see an example in \Cref{fig:entropy}~\cite{torres-gomez2020teaching}.
Video sequences that occur in peaceful environments where elements are slow-moving will present a lower level of entropy than those with higher variability.
For example, with the video sequence ``500 Years of Women in Western Art'' by Philip Scott Johnson~\cite{noauthor_500_nodate}, students are asked to evaluate the random evolution of entropy in the Renaissance period and assess more deterministic transition in the Cubism style for female portraits.

%

\subsubsection{Assessing the practice of arts}

When evaluating the practice with students, we applied quantitative and qualitative questionnaires.
In the quantitative questionnaire, we graded the student's learning about specific concepts and their technical skills with a final exam (more details in~\cite{torres-gomez2021teaching}).
Our results show quite an improvement in the scores with the last five editions: an average of \num{56.64},\footnote{All the scores are over \num{100}.} over the \num{23} students taking the examination versus a mean of \num{47.54} with \num{20.20} students macro-averaged over the previous five years.\footnote{Past edition 5 seems to be an outlier since a different teacher with different evaluation criteria was in charge of the course.}

\begin{figure}
    \centering
    \includegraphics[width=0.7\columnwidth]{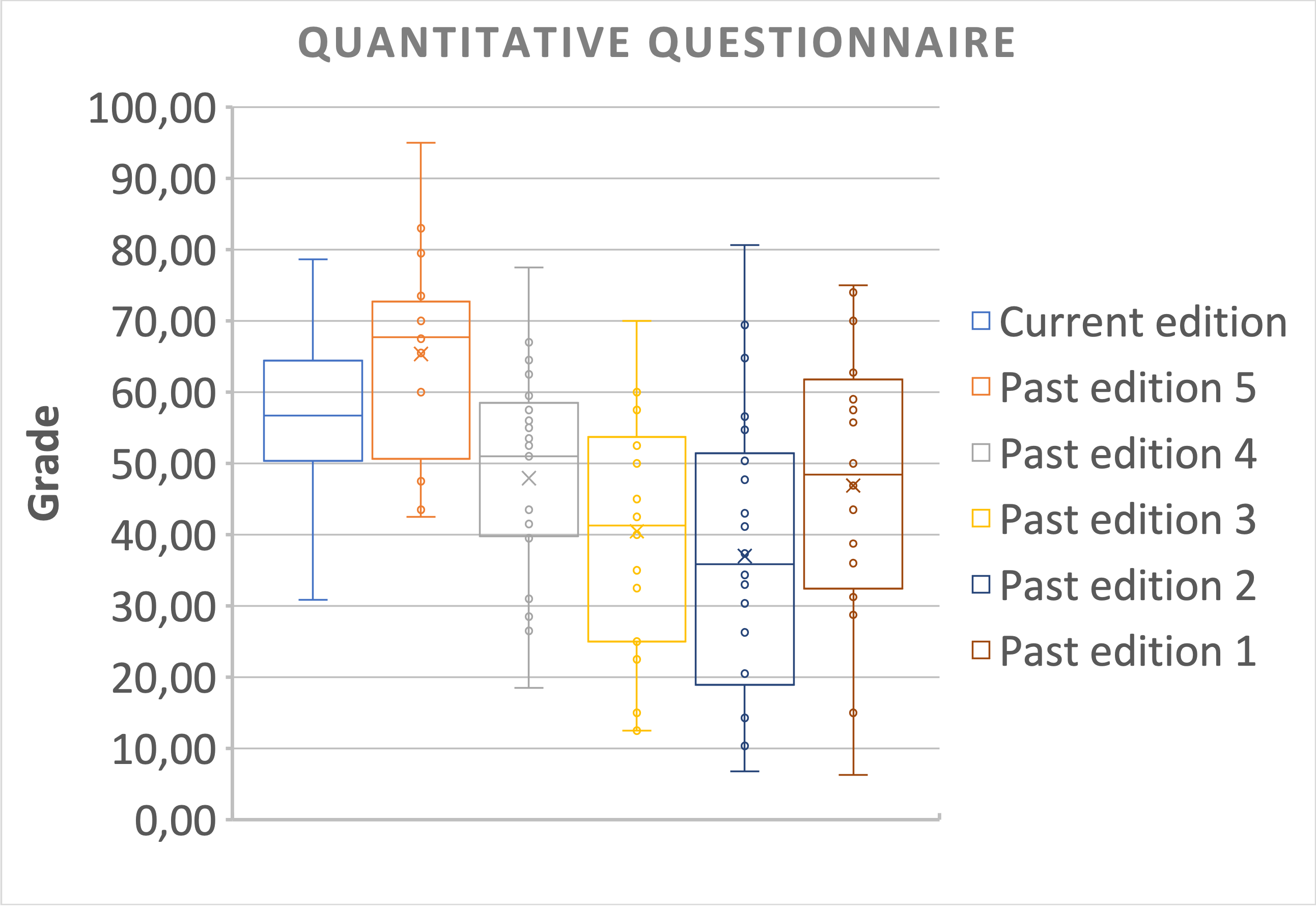}
    \caption{Comparative results for the quantitative questionnaire.}
\label{fig:quantitave_questionnaire}
\end{figure}

\begin{figure}
    \centering
    \includegraphics[width=0.7\columnwidth]{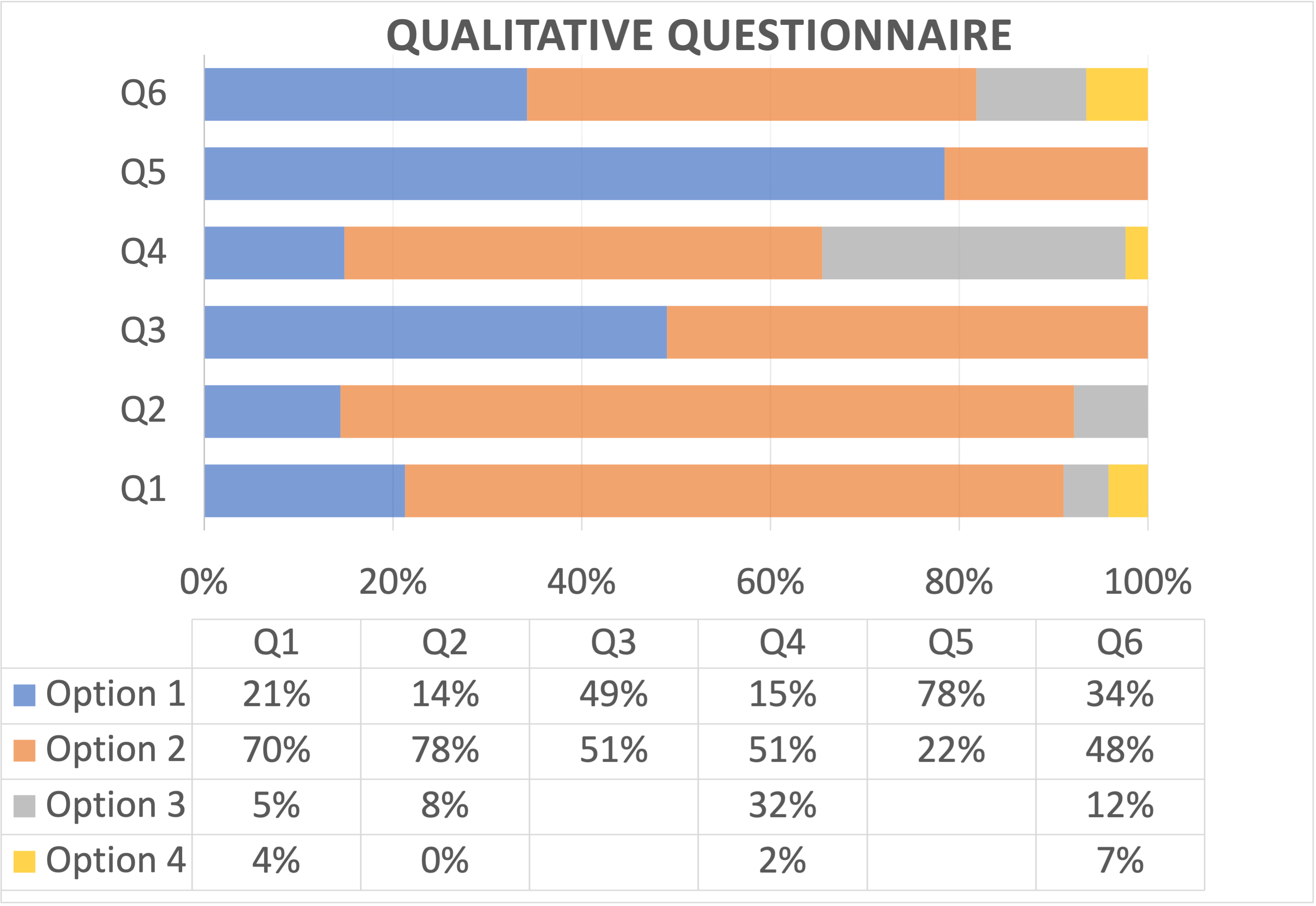}
    \caption{Comparative results for the qualitative questionnaire (ordinary call).}
\label{fig:qualitave_questionnaire}
\end{figure}

We applied a qualitative questionnaire (see \Cref{fig:qualitave_questionnaire}) consisting of six questions and various options per question taken four times.
An average of \num{13.33} students answered these surveys to the following questions.
Q1) \textit{The activity was (very, enough, barely, not)} interesting; Q2) \textit{The activity was (very, barely, not)} complex; Q3) \textit{My previous knowledge of the contents was (enough, not enough) to complete the lab exercise}; Q4) \textit{The information provided for the lab exercise was (very adequate, adequate, scarce, very scarce)}; Q5) \textit{ The explanations of the instructor (greatly, barely)} helped me; and Q6) \textit{The lab exercise helped me (very much, somewhat, a few, nothing) to understand key concepts}. 
From this plot, we can conclude that most of the students found the exercises interesting and felt that they helped them understand key targeted concepts.
However, they also found the exercises somewhat complex (because of the efforts required to derive meaningful results), and some of them would have liked further guidance material.

%

\subsection{Oxford-Style: Performing debates in classrooms}

We can also consider a different approach, applying the Oxford-style debate, as a means to develop the teaching of professional and technical skills jointly.  The Oxford-style debate is a formal communication process where participants argue for and against a predetermined topic, called motion.

\begin{figure}
\centering
\includegraphics[trim= 0 70mm 230mm 0, clip,width=0.3 \textwidth]{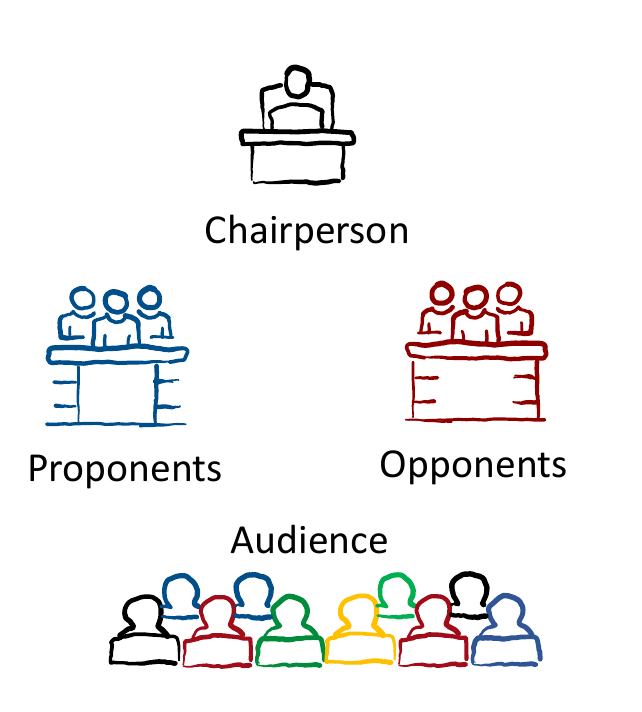} 
\caption{Oxford-style debate scheme.}
\label{fig:oxf}
\end{figure}

To illustrate, we develop the topic '\textit{Opening the source code increases the security level of the application}', allowing us to elaborate on various technical concepts (following the \textit{Step~\num{1}} of the methodology).
The discussion follows a competitive format in which participants are divided into two teams: proponents, who support the motion, and opponents, who argue against it; see an example in \Cref{fig:oxf}. 
Usually, each group consists of four students, and each of them has been assigned a specific role, introducing to the discussion, presenting arguments and counterarguments, etc., which also allows students to prepare scripts and perform like in a theatre (\textit{Step~\num{2}} as the process is inspired by performing arts. 

Speakers from both sides take the floor alternately. 
Each speaker has a strict limit of time -- usually \num{5} minutes -- to convince the audience of the motion or against the motion. 
The debate is led by the chairperson (usually the teacher), who supervises the course of the discussion, announces the speakers, gives the floor, and also has the right to take the floor (the crucial \textit{Step~\num{3}}). 
After the debate, the audience (the rest of the class) assesses the debate and provides the result. 
Constructive feedback for individual participants in the debate is important because the debate has an educational purpose (assessment as \textit{Step~\num{4}}).

An Oxford-style debate is based on factual arguments. 
Therefore, the participants improve their professional knowledge. 
We remark that all speakers must thoroughly investigate the topic and identify both pro and con arguments to prepare for their speech and for possible questions during the debate. 
Such a rigorous self-study emphasizes critical thinking of one's own opinion. 
Therefore, participants, including the audience, take a broad view of the given topic, fully aware of the existence of different points of view.
Furthermore, debate is a powerful tool for efficient teamwork, a crucial skill for modern ICT engineers~\cite{mn-2021}.
All team members must cooperate to prepare a coherent line of argumentation.  

Oxford-style debate enhances participants' professional skills, particularly in communication, social skills, and emotional intelligence. 
These skills are essential for engineers and future entrepreneurs who use them in everyday business practice. 
Speakers also learn the ability of non-verbal communication: correct diction, voice modulation, clear articulation, convincing body language, gesticulation, eye contact with the audience, etc. 
Such techniques help engineers improve communication effectiveness and energise public presentations (i.e., during business meetings in the future). 
Active participation in debates develops negotiation and persuasion skills. 
In addition, students are trained to manage and reduce their stress level during public speeches.

%

\subsubsection{Assessing the Oxford-style from students}
ICT engineers who actively participated in Oxford-style debates appreciate this approach to teaching. From 2020 to 2022, students completed anonymous surveys after participating in Oxford-style debates~\cite{niemiec_2022}. 
We also asked participants to rate this didactic tool as very useful: score~\num{9} on a scale from~\num{0} (completely useless) to~\num{10} (very useful). It is worth mentioning that the participants in Oxford-style debates passed the course in the first evaluation round.

\begin{figure}
    \centering
    \includegraphics[width=0.7\columnwidth]{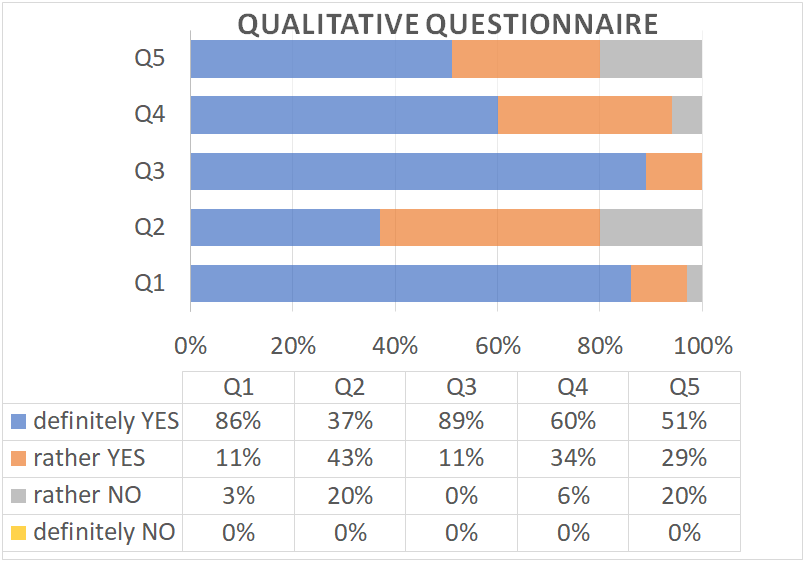}
    \caption{Qualitative assessment of Oxford-style debates.}
\label{fig:qual_questionnaire_debates}
\end{figure}

The results of the qualitative assessment of the Oxford-style debate are presented in \Cref{fig:qual_questionnaire_debates}. 
It consists of five questions and four different options per each question: \textit{definitely YES}, \textit{rather YES}, \textit{rather NO}, or \textit{more definitive NO}.
An average of \num{12} students per year answered the survey to the following questions.
Q1) \textit{Has participation in the debates improved your professional skills?}; Q2) \textit{Has participation in the debates improved your technology skills/IT knowledge?}; Q3) \textit{Do you think participating in the debate as a speaker was a valuable experience for you?}; Q4) \textit{Do you think participating in the debate as a judge was a valuable experience for you?}; and Q5) \textit{Do you think participating in the debates made you a better IT engineer?} 
The attendees confirmed that the Oxford-style debates significantly improved their professional and technical skills. 
The students acknowledged both roles as valuable; however, participation as an active speaker is assessed more highly than as a judge. 
Finally, the vast majority confirmed that participating in Oxford-style debates improved their skills as \ac{ICT} engineers.

%

\subsection{Gamification}

Gamification is the use of game elements and game design techniques in a context not related to games, aimed at engaging people, motivating them to action, stimulating them to learn, and solving problems while achieving the desired learning outcomes~\cite{DBLP:conf/mindtrek/DeterdingDKN11, DBLP:journals/electronicmarkets/HuotariH17}.
Gamification can be understood as a product way of thinking, process, experience, and approach to design elements of mechanics known from games (e.g., computer games).
In education, the goal is to improve the degree of learning and the level of involvement with the content of the course for the students~\cite{caponetto2014gamification, DBLP:conf/sigcse/IosupE14}.
This section illustrates our experience with two different approaches applied at different locations.

The social gamification, as follows from the work in~\cite{DBLP:conf/sigcse/IosupE14} focusses on redesigning courses as games with a strong social component. This approach addresses the context in which students are largely free to pursue a profession of their choice, while the teaching staff has limited resources to manage large, diverse classrooms across both the Bachelor's and Master's cycles. In this context, social gamification aims to help students discover their inner motivation and can help persuade the undecided to take educational steps in the right direction.  
This approach has been refined over many years in several computer systems courses taught by different (but not entirely independent) educators, and tested at a technical and research university in the Netherlands.
\par For example, in the first-year BSc course on Computer Organisation, which is a common first step in the computer systems part of all computer science and engineering curriculum,\footnote{In particular, the ACM/IEEE CC 2020, CS 2013, and CE 2016.} the course designers decided to enrich the entire content~(\textit{Step 1}). 
Through a holistic approach, the designers added social gamification elements to the course to support motivation and participation~(\textit{Step 2}). 
The Computer Organisation course addresses \textit{Step 3} 
with specific activities, learning objects, and measures. Students can choose their path of advancement and depth of knowledge and can choose to engage more in content designed to stimulate exploration, social engagement, competition (player vs. player, PvP), or achievement (player vs. environment, PvE). This contrasts with traditional approaches that limit choice, cater to a single prototypical student, and/or try to spur only one kind of behaviour (e.g., only competition or only achievement measured against content milestones).

An important element is the management of course dynamics~(part of Step~\num{3}, continuously evaluated in Step~\num{4}). 
Various types of student-to-student and student-to-classroom interactions can be important for social gamification. Examples in the Computer Organisation course include presentation of own projects, competition around a specific learning point in the lecture or at the end of project development, help with the exploration of specific advanced topics, extra lectures for the group of best students, etc. 

As an another example, we developed an evaluation system where the final grade is computed ensuring that each participating member's contribution is accounted for.
To complete the subject, students must perform well on the most important technical topics related to image processing, such as logical and arithmetical operations on images, mathematical morphological operations on images, digital image filters, etc.; all of these topics are part of the \textit{Step~1} of our methodology.
To implement this approach, each participant member is represented as a pixel in the multimedia world, aiming at producing a full-fledged, breathtaking video as the final goal (\textit{Step~2}).
The Pixel will be transformed into greyscale and then into a Wide Colour Gamut (WCG) according to the collected results as: ``Knowledge points'' by attending the talk and marking one's attendance, ``practice points'' (by passing exercise lessons/assignments), ``Skill points'' (by off-line works) and ``Test points'' by quiz (\textit{Step~4}). 
The collected {K}nowledge, {P}ractice, {S}kill, and {T}est points come up with an absolute value higher than the required maximum, so students can trade activities with other ones (to some extent, to avoid full substitution of one activity for another). 
The sum of the collected points converts to the attribute and then to the final grade, according to the official regulations (so the gamification is compliant with regular teaching rules).

%

\subsubsection{Assessing Gamification in Classrooms}

\begin{figure}
\centering
\includegraphics[width=0.7\columnwidth]{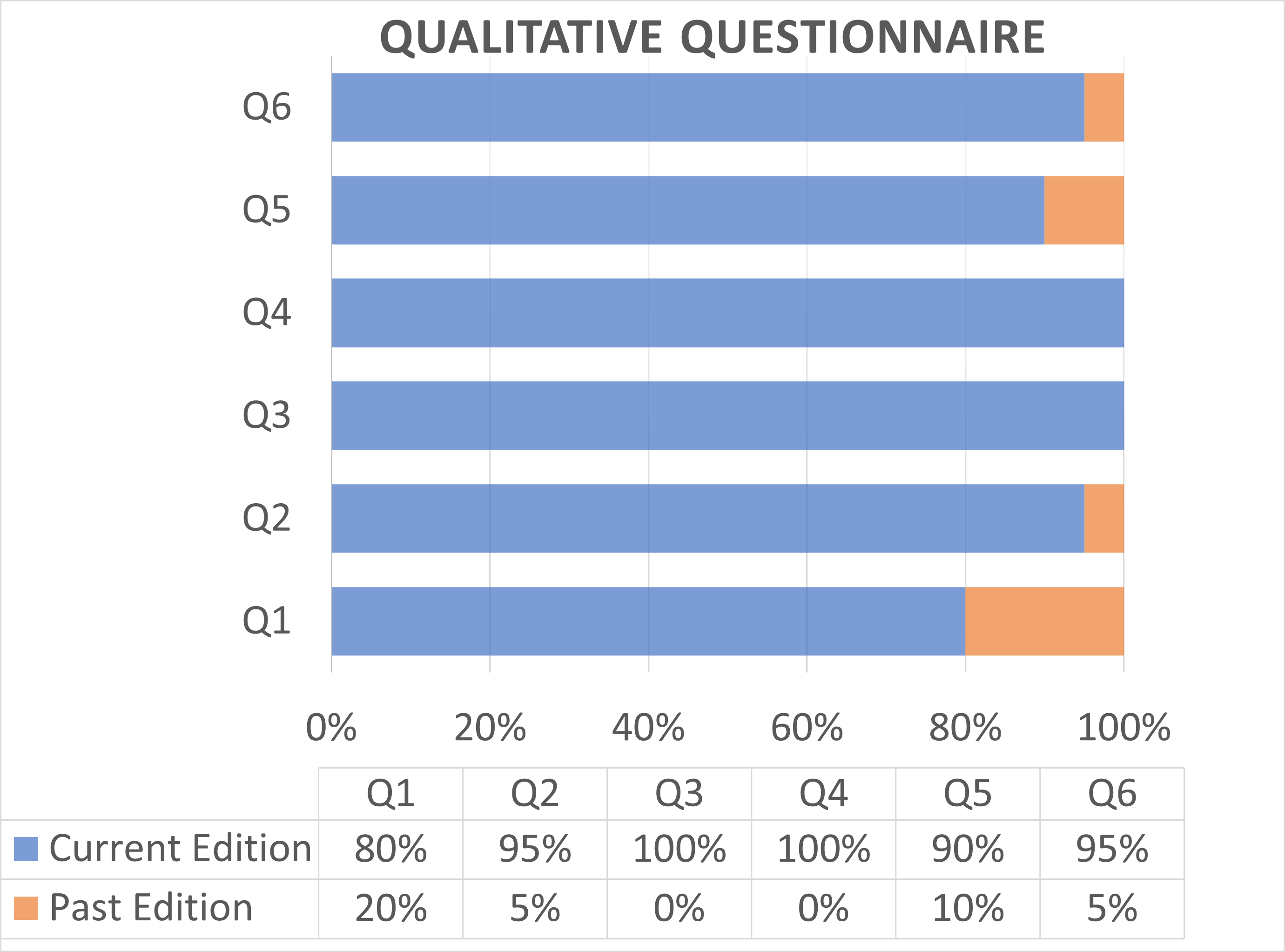}
\caption{Results of the evaluation of new teaching methods with the gamification technique.}
\label{fig:evaluation}
\end{figure}

An important step is evaluating the student's knowledge and perception (\textit{Step~4}). 
Using a social gamification approach, we report that more students completed the course and engaged with advanced topics than in previous editions without gamification. 
To illustrate the results compared to methods previously used, an evaluation was carried out on a group of \num{53} students, of whom \num{18} responded to the surveys; see a summary in \Cref{fig:evaluation}. 
The evaluation of new methods was carried out in six categories: adequacy to the needs of recipients (Q1), effectiveness in achieving goals/results (Q2), usefulness of the results for recipients (Q3), efficiency, i.e., the ratio of effort to benefits (Q4), the permanence of the results (Q5), and impact on students and their environment (Q6). 
The exercise classification technique was used, and in each case, the students indicated their preference for the new methods over the ones previously used.
Taking into account the approach applied at AGH University, we confirmed increased interest and participation of students during classes after evaluations and the development of individual practices with high attendance rates.

%

\section{Steps Forward}
\label{sec:discussion}

The methodology and examples presented above illustrate specific actions for converging the two schools. Advancing its applicability, we identify further remarks and open issues concerning engagement from teachers, motivation for females to follow STEM studies, and scalability of such proposals.

\textbf{Remark A: Improved engagement of teachers.} We posit that integration of arts-based concepts and methods with \ac{ICT} subjects also benefits teachers.
Educators have a good opportunity to build or improve good relationships with students and extend their own professional knowledge. 
Among others, motivated students often enrich discussions and exercises with relevant examples from their personal experiences. 
Therefore, sharing knowledge in a group can be another important motivation for a teacher.

\textbf{Remark B: Expediency to promote \ac{STEM} careers at preuniversity levels.} 
\ac{STEM} schools are usually based on hands-on activities and logical and problem solving-orientated approaches. 
The liberal arts might be an approach that will help students gain \ac{STEM} skills through an alternative approach that will allow them to grow their level of understanding of physical processes (engineering) in technical subjects and increase the motivation of students. 
The liberal arts approach in technical universities will help young people (not only from vocational, professional, or technical schools) get in touch with \ac{STEM} requirements during technical university visits. 
That will also positively influence the teachers accompanying the students during the visits. 

\textbf{Remark C: Extension to other fields of engineering study}.
This approach might be extended to any engineering field. 
For instance, in civil engineering, a variety of concepts; including construction materials, technologies, and management, are typically taught in their own theoretical silos.
Typically, there is a lack of means to integrate the skills and abilities of all the people involved when working with projects in practice.
In this sense, the ability to view, analyse, and understand different disciplines from a comprehensive point of view can be supported by the methodologies of the liberal arts.

\textbf{Open Issue A: Scalability of such methodologies.}
To successfully implement this concept, some conditions must be met. 
On the teacher's side, it is crucial that she/he is convinced of the importance of professional skills and that it is a valuable teaching and learning objective. 
On the curriculum side, there is the necessity for 'space' and preparation for such teaching objects. 
However, guidelines and best practices are generally not available to naturally include the teaching of professional skills in technical subjects as a single unit.
The above-presented methodology and examples illustrate its affordability in a variety of subjects; however, its generalisation only comes after intentional generalisation practices, where the role of institutional resorts plays a relevant matter. 

\textbf{Open Issue B: Reinforcing females participation in \ac{STEM} careers.}
Many questions arise when considering the potential benefits of including arts in \ac{STEM} education as a means of filling the gender gap existing in \ac{STEM}-related higher education careers. 
According to Eurostat \cite{eurostatEdu2018}, the ratio between male and female engineering students is $2.53$ while the reverse situation appears in the arts and humanities where the ratio of women vs. men is \num{2.09}. 
In fact, gender stereotypes are often mentioned in the literature~\cite{makarovaGenderGapSTEM2019,garcia-holgadoGenderEqualitySTEM2020} as having a great impact on student career aspirations. 
In this regard, a research question left open in this paper is whether using arts as a vehicle to understand STEM topics could contribute to changing the male stereotypes underlying STEM subjects and, therefore, improving women's engagement. 

\textbf{Open Issue C: Adequate the methodology in practice.}
We identify a challenge related to putting into practice the methodology introduced in \Cref{sec:methods}.
Because many fields in \ac{STEM} have a specific broader culture and some have spurred the emergence of community and localised practices, adapting the general methodology to local conditions becomes important and is a nontrivial task. 

We also identify an opportunity: by reframing the problem of change into one of observing art seen as a process, notably, an \textit{external} process, the attitudes of both students and teachers could change more easily than if the change would be framed in local, technical terms.

\textbf{Open Issue D: Qualitative vs. Quantitative Evaluations}
Additionally, a serious introduction to our proposed concepts requires an adequate evaluation practice. 
In this case, due to the teaching content (i.e., professional skills), qualitative (instead of quantitative) evaluations for this specific teaching concept need to be designed. 
It is still challenging to design suitable evaluation methods, as a balance between qualitative and quantitative.

%

\section{Conclusion}
\label{sec:conclusion}

This contribution provides a pragmatic framework of general applicability for teaching together professional and technical skills in technical careers.
From the student's perspective, the intertwining of the liberal arts approaches with technical exercises reflects a contributing practice to the acquisition of professional and technical skills.
However, the effective integration of the two schools is not still mature in practice, although more than sixty years have elapsed since Snow's speech. 
More efforts are needed to develop guidelines for the wider applicability of conducting practices to be applied in classrooms.

%

\bibliography{references}

\end{document}